\documentclass{rmaa}

\title{Compact radio sources in the vicinity of the ultracompact HII region G78.4+2.6}

 \author{Citlali Neria, Yolanda G\'omez, and Luis F. Rodr\'\i guez 
  \affil{Centro de Radioastronom\'{\i}a y Astrof\'{\i}sica, UNAM, Morelia}
 }

\fulladdresses{
\item Citlali Neria, Yolanda G\'omez,
Luis F. Rodr\'\i guez: Centro de Radioastronom\'{\i}a
y Astrof\'\i sica, 
UNAM, A. P. 3-72, 
(Xangari), 58089 Morelia, Michoac\'an, M\'exico
(c.neria, y.gomez, l.rodriguez@crya.unam.mx).
}

\shortauthor{Neria, G\'omez \& Rodr\'\i guez}
\shorttitle{Compact radio sources in the vicinity of G78.4+2.6}

\SetVolume{00} \SetFirstPage{001} \SetYear{2010}
\ReceivedDate{\today} 
\AcceptedDate{Year Month Day} 

\resumen{Utilizando el Very Large Array (VLA) a 3.6~cm,
hemos identificado cuatro nuevas fuentes compactas de radio 
en los alrededores de la regi\'on HII con morfolog\'i{}a 
cometaria G78.4+2.6 (VLA~1).
Las cuatro fuentes compactas de radio (denominadas VLA~2 a 5), 
tienen contrapartes en el infrarrojo cercano, como se muestra en 
la imagen del Spitzer a 3.6 $\mu$m. Una de estas fuentes (VLA~5) claramente 
muestra evidencia de radio-variabilidad en una escala de tiempo de horas.
Hemos explorado la posibilidad de que estas radio-fuentes  est\'an 
asociadas con 
estrellas de pre-secuencia principal (PMS) en la vecindad de 
la regi\'on HII ultracompacta G78.4+2.6. Nuestros resultados
favorecen una distancia corta de 1.7~kpc para G78.4+2.6. 
Adem\'as de la detecci\'on de radio-fuentes en los alrededores 
de G78.4+2.6, detectamos otro grupo de cinco fuentes, localizado 
aproximadamente 3$^{'}$ al noroeste de la regi\'on HII. Algunas de estas 
fuentes presentan emisi\'on extendida.}

\abstract{Using the Very Large Array (VLA) at 3.6~cm
we identify four new compact radio sources 
in the vicinity of the cometary HII region G78.4+2.6 (VLA~1). 
The four compact radio sources (named VLA~2 to VLA~5), 
have near-infrared counterparts, as seen in the 3.6 $\mu$m 
Spitzer image. One of them (VLA~5) 
clearly shows evidence of 
radio variability in a timescale of hours. 
We explore the possibility that these radio sources are
associated with pre-main sequence (PMS) stars in the vicinity of 
the UC HII region G78.4+2.6. Our results 
favor the smaller distance value of 1.7 kpc for G78.4+2.6. 
In addition to the detection of the radio sources 
in the vicinity of G78.4+2.6, we detected another 
group of five sources which appear located about 3$^{'}$ to the 
northwest of the HII region. Some of them exhibit extended emission.}


\begin{document}

\maketitle

\section{Introduction}
Although our understanding of star forming regions has progressed in recent 
years, many aspects of the early evolution stages are still unclear. 
Observationally, pre-main sequence (PMS) stars are among the most 
difficult to study because they are usually heavily extincted by their 
parental molecular cloud. Observations at radio wavelengths of young 
clusters are of considerable relevance because a fraction of these 
stars appear to be undetectable even in the infrared. In this sense 
radio observations can provide unique information. Radio clusters of 
non-thermal PMS stars have been found in several star forming 
regions. G\'omez et al. (2000; 2002) reported centimeter wavelength 
observations of the GGD14 star forming region. In this study, they 
found a radio cluster of PMS stars made of eight extremely 
compact and faint  radio sources around a compact HII region, that
seems to be ionized by a B0.5 zero age main sequence (ZAMS) star. 
Similar radio clusters 
have been studied toward HH~124~IRS by Reipurth et al. (2002),
NGC~2024 (Orion B) by Rodr\'{i}guez et al. (2003) 
and toward the Orion Nebula (M42) by Zapata et al. (2004), among others.
Some members of the radio clusters mentioned above are characterized 
by compact non-thermal emission, time variability, and in some cases 
circular polarization. All these characteristics suggest that most 
likely their radio emission is due to the gyrosynchrotron mechanism 
from an active magnetosphere in a PMS star (e.g. G\'omez et al. 2002).

The identification of more such PMS clusters can provide valuable 
information of the star forming processes. Here we present 
the identification of new members that surround the HII 
region G78.4+2.6. This is an ultracompact (UC) HII region where
estimated kinematic distances go from 1.7 kpc (Wilking et al. 1989;
Schneider et al. 2006) to 3.3~kpc (Kurtz et al. 1994).
It has a cometary morphology (Kurtz et al. 1994) and is associated with 
the IRAS source 20178+4046 with a far-infrared luminosity of 
7$\times$10$^4$ L$_{\hbox{$\odot$}}$ (assuming a distance of 3.3 kpc). 
From the radio continuum emission the HII region seems to be ionized 
by a B0$-$B0.5 ZAMS star (Kurtz et al. 1994, Tej et al. 2007). 
No maser emission of OH (6035 MHz), $\mathsf{H}_2$O (22.2 GHz) or 
CH$_3$OH (6.7 GHz) or thermal emission of NH$_3$ (23.6-23.9 GHz) has been 
detected toward this region (Baudry et al. 1997; Codella et al. 1996; 
Kurtz \& Hofner 2005; Sunada et al. 2007; Slysh et al. 1999). 
In the mid-infrared images the UC HII region appears single 
(Crowther \& Conti, 2003), but the presence of a stellar 
group/cluster around it is clear in the 
near-infrared images (Tej et al. 2007). 

Recently, Tej et al. (2007) presented a multiwavelength study toward 
IRAS~20178+4046. They used the Giant Metrewave Radio Telescope 
(GMRT) to map the radio continuum emission at 1280 and 610 MHz. 
Available archival data from the Two Micron All Sky Survey (2MASS), 
the Midcourse Space Experiment (MSX), and the James Clerk Maxwell 
Telescope (JCMT) were used to study the complex in the near-infrared (NIR)
JHKs bands, the mid-infrared at 8.3, 12.13, 14.65 and 21.34 $\mu$m and the
sub-mm at 450 and 850 $\mu$m, respectively.
From their analysis of the NIR 2MASS data, they
proposed the presence of 
nine early type sources with spectral types $\sim$B0.5 or earlier around 
IRAS~20178+4046. Their multiwavelength 
study led Tej et al. (2007) to conclude that an scenario, 
with possible different evolutionary stages in star formation 
is present toward IRAS~20178+4046 (G78.4+2.6).  

In this paper we present sensitive high resolution radio 
continuum observations at 
3.6~cm, in order to search for faint compact radio continuum sources 
around this UC HII region. 

\begin{figure}
\centering
\includegraphics[scale=0.45, angle=0]{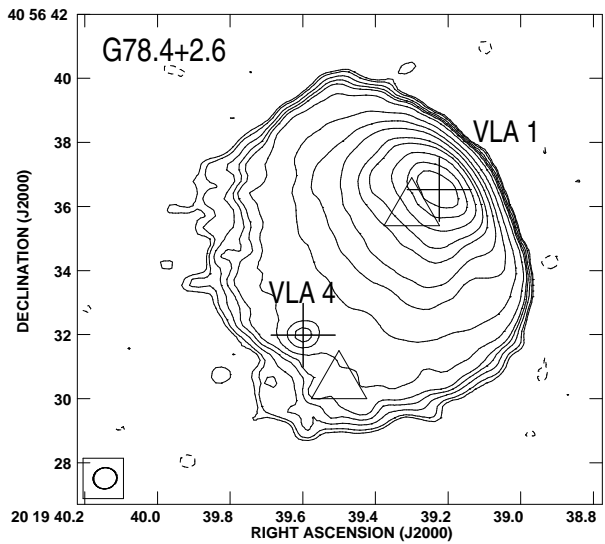}
\caption{Contour image of the 3.6 cm continuum
emission from the UC HII region G78.4+2.6.
The contours are -3, 3, 4, 6, 8, 10, 15, 30, 50, 70, 100, 120, 150, 180 and 200
times 10~$\mu$Jy~beam$^{-1}$,
the rms noise of the image.
The synthesized beam  ($0\rlap.{''}76 \times 0\rlap.{''}67$
with a position angle of $-$79$^\circ$) is shown in the bottom left corner
of the image. The dynamic range achieved in the 3.6~cm image is approximately
200. The crosses mark the peak position of the 3.6~cm radio sources VLA~1
and VLA~4. The triangles mark the positions of the NIR sources number 1 (bottom
left) and number 2 (top right) of 
Tej et al. (2007).}
\label{fig1}
\end{figure}

\section{Observations}
G78.4+2.6 was observed with the VLA of the NRAO\footnote{The National Radio 
Astronomy Observatory (NRAO) is operated by Associated Universities 
Inc. under cooperative agreement with the National Science Foundation.} 
at 3.6~cm (8.4~GHz) in the B configuration during 2002 August 6 and 7
for a total on-source time of $\sim$10 hrs. 
The data were taken using two IFs each with an effective bandwidth of 50~MHz
and both circular polarizations. 
The flux density scale was determined from observations of the amplitude 
calibrator 1331+305, for which we assumed a flux density of 5.22 and 5.20 Jy, 
for IF=1 (8.4351 GHz) and IF=2 (8.4851 GHz), respectively. The source 
2007+404 was used as the phase calibrator, with bootstrapped flux densities of 
2.704$\pm$0.003 and 2.708$\pm$0.003 Jy for the first and second IF, 
respectively. The data were edited and calibrated using the standard
procedures of the Astronomical Image Processing System (AIPS) of NRAO. 
The data were also self-calibrated in phase and for sources far away 
from the center we applied a primary beam correction to the images 
(with the task PBCOR). The synthesized beam is 
$0\rlap.{''}76 \times 0\rlap.{''}67$ with a position angle of 
$-$79$^\circ$, and the $rms$ noise is 10~$\mu$Jy~beam$^{-1}$.\\

\section{Results}
 \subsection{Compact radio sources toward G78.4+2.6}

Figures~1 and 2 show the radio continuum emission at 3.6~cm toward the UC HII 
region G78.4+2.6. In addition to the UC HII region, which we hereafter 
call VLA~1, it is possible to identify emission from four fainter 
compact radio continuum 
sources above the $5\sigma$ level, they are called VLA~2 through VLA~5 by us.
Figure~1 shows VLA~1 and the compact source VLA~4, which appears inside 
the extended emission region of the cometary source VLA~1. 
The source VLA~4 could be physically embedded in the cometary HII
region. Unfortunately, our data do not have enough angular
resolution to ascertain the nature of this radio source. 
Figure~2 exhibits a larger area where all the compact radio sources 
in the vicinity of VLA~1 can be identified.

\begin{figure*}
\centering
\includegraphics[scale=0.8, angle=0]{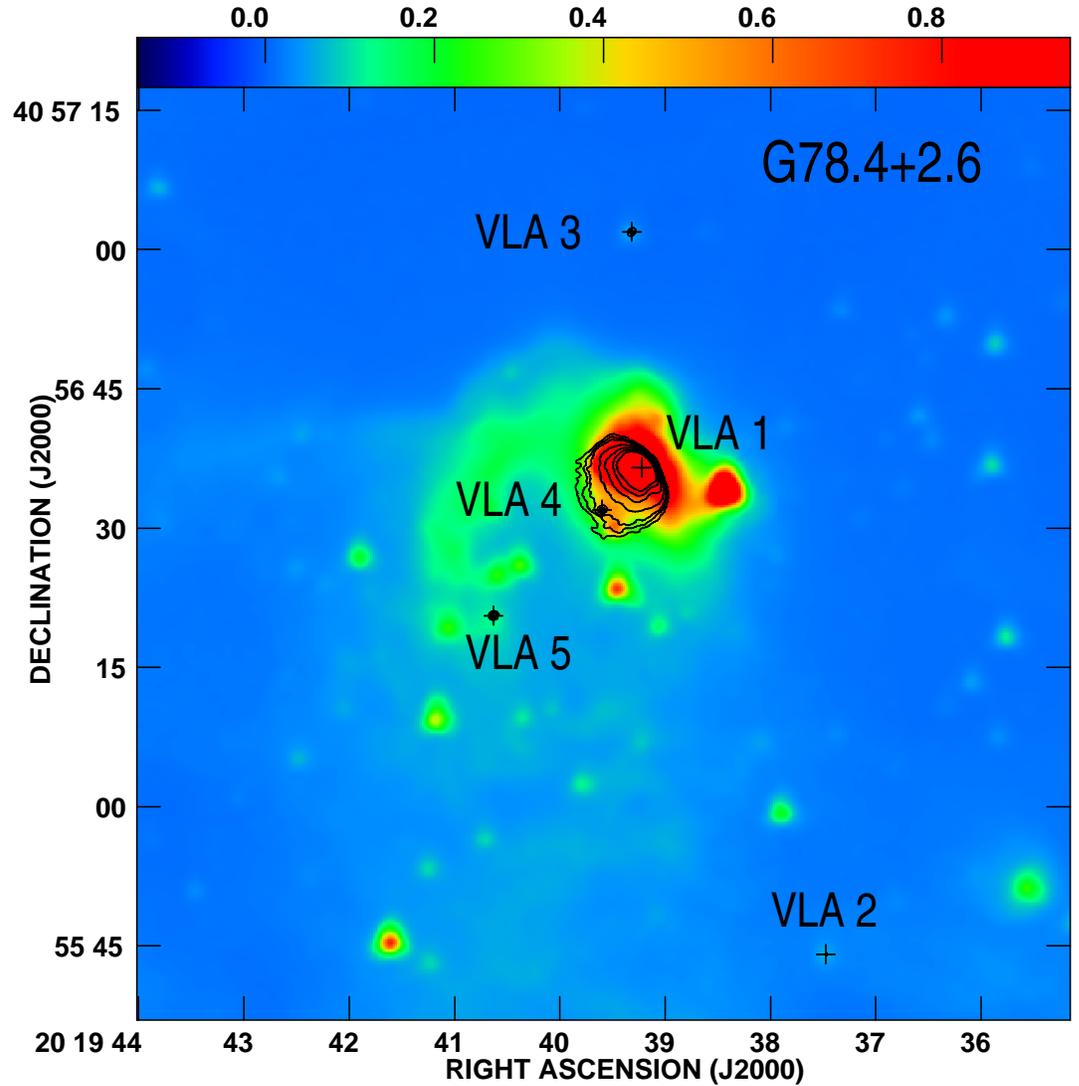}
\caption{The image shows the Spitzer 3.6 $\mu$m band emission
in color for the region around G78.4+2.6. Color
scale goes from $-$0.15 to 0.95 GJy~sr$^{-1}$.
The contours represent the 3.6 cm continuum
emission from G78.4+2.6 showing the four compact radio sources.
The contours are -5, 5, 10, 15, 30, 50, 70 and 90 times 10 $\mu$Jy~beam$^{-1}$,
the average $rms$ noise of the image. 
The crosses mark the peak position of the 3.6 cm radio sources listed in
Table 1.  The synthesized beam is
$0\rlap.{''}76 \times 0\rlap.{''}67$ with a position angle of $-$79$^\circ$.
}
\label{fig2}
\end{figure*}
 
For these compact sources, we looked for near-infrared counterparts 
using the Spitzer 3.6~$\mu$m band image and the 
2MASS Point Source Catalog (PSC),
within a radius of 2$''$ around each VLA source.
The parameters of the radio continuum sources, and their possible 
counterparts, are listed in Table~1.
Of the nine NIR 2MASS sources reported by Tej et al. (2007), 
only one (source 2) clearly coincides in position with a 
VLA source (VLA~1). There is another NIR source reported by 
Tej (source number 1)
that could be associated with a VLA source (VLA~4; within 
1$\rlap.{''}$5). In Figure~1 we plot the peak position of the NIR source
number 1 of Tej et al. (2007), where the shift in position
with respect to the radio source VLA~4 is evident. Further 
observations at different frequencies 
and higher angular resolution are required to confirm or reject the 
association. Tej et al. (2007) have noted that their NIR source 
number 1 coincides in position with the source IRAS~20178+4046, one
of the brightest infrared sources in the region.

To study the time-variability of the four fainter compact radio continuum 
sources (see Figure~2) during the approximately 12 hours
of the total observing session (of which 10 hours were on-source), 
we proceeded as follows.
The precise position of each source was determined from the image
shown in Figure 2. The \sl (u,v) \rm data was then recentered at the position
of the source considered and the real and imaginary parts of its
flux density were plotted as a function of time, averaged 
over the \sl (u,v) \rm plane and in 10 time bins.
A detailed description of this technique is given in the Appendix.
To avoid contamination from extended emission, we use only \sl (u,v) \rm data
with baselines larger than 100 k$\lambda$, suppressing emission from
scales larger than $\sim 0\rlap.{''}5$. The technique did not work in a
reliable way in the case of VLA 4, that is embedded in the much
brighter UC HII (VLA 1). In the case of the remaining sources, we found
clear evidence of variability for VLA 5. In Figure 3 we show the real 
and imaginary parts of its
flux density as a function of time, where it can be 
appreciated that the source
was very weak during the first half of the run and that, in a timescale of
about an hour, it rose to a much stronger flux density. No evidence of
variability was found for VLA 2 and VLA 3.
In Table~1, column (8) we listed if the sources were time-variable or not. 

\begin{figure}
\centering
\includegraphics[scale=0.4, angle=0]{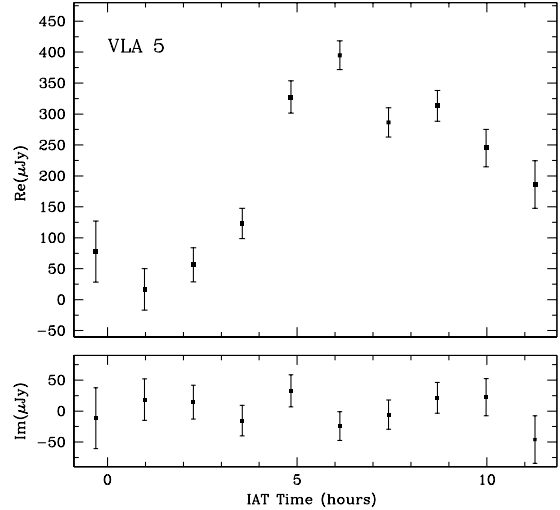}
\caption{Real (top) and imaginary (bottom) components of
the flux density of VLA 5 as a function of
time (IAT = International Atomic Time).
}
\label{fig3}
\end{figure}

The variability observed in VLA 5 has been observed in other
low-mass PMS stars (e. g. G\'omez et al. 2002; Bower et al. 2003;
G\'omez et al. 2008; Forbrich et al. 2008) and 
it can be attributed to 
gyrosynchrotron emission from an active magnetosphere.
For UC~HII regions or externally-ionized globules variability
at such short time scales is not expected.
  
\begin{table*}[htbp]\centering
\small
  \setlength{\tabnotewidth}{2.0\columnwidth} 
  \tablecols{8}
  \caption{Parameters of the radio sources toward G78.4+2.6}
 \begin{tabular}{c c c c c c c c}
 \toprule
 Source &$\alpha$(2000) &$\delta$(2000) &Flux Density\tabnotemark{a} &NIR Counterpart &\multicolumn{2}{c}  {Spitzer Counterpart\tabnotemark{b}} &Time\\ 
  & & &(mJy)& (2MASS~PSC)& $\alpha$(2000)&$\delta$(2000) &Variable?\\
 (1)&(2)&(3)&(4)&(5)&(6)&(7)&(8)\\
 \midrule
VLA~1 &20 19 39.224  &40 56 36.53 &62.26$\pm$0.14 &20193932+4056358\tabnotemark{d} &20 19 39.12 &40 56 37.84 &N\\
VLA~2 &20 19 37.475  &40 55 44.11 &0.06$\pm$0.02 &20193748+4055441 &20 19 37.48  &40 55 44.14 &N\\
VLA~3 &20 19 39.318  &40 57 01.91 &0.11$\pm$0.02 &20193931+4057018  &20 19 39.34 &40 57 01.85 &N\\
VLA~4 &20 19 39.599  &40 56 31.98 &0.48$\pm$0.02 &20193949+4056305\tabnotemark{e} &20 19 39.56 &40 56 32.18 &N\\
VLA~5 &20 19 40.630  &40 56 20.57 &0.24$\pm$0.02\tabnotemark{c} &-- &20 19 40.64 &40 56 20.19 &Y\\
 \bottomrule
 \tabnotetext{a}{Flux densities at 3.6~cm. The flux densities for sources 
VLA~2 to 5 were measured from an image where extended emission was suppressed 
using only ($u$, $v$) data with baselines larger than 100~k$\lambda$).}
 \tabnotetext{b}{Peak position associated with the 3.6~$\mu$m Spitzer image.}
 \tabnotetext{c}{Time-averaged flux density}
 \tabnotetext{d}{NIR source number 2 of Tej et al. (2007)}
\tabnotetext{e}{NIR source number 1 of Tej et al. (2007)}
 \label{tab:1}
 \end{tabular}
\end{table*}


\subsection{Other radio sources in the field}
Toward the north of the UC~HII region G78.4+2.6, we detect
another cluster of five compact radio sources (designated as VLAN~1
to VLAN~5) above $5\sigma$ level. After we
applied UVTAPER (50 k$\lambda$, resulting in a beam size of 
3$\rlap.{''}$26$\times$3$\rlap.{''}$02, P.A.=$-$60$^\circ$) to 
this image we noted
that two of these sources (VLAN~2 and 3) have diffuse 
extended emission associated with them. The extended radio 
emission toward the east of VLAN~2 is more likely an extension of a 
possible jet. VLAN~1 also displays some 
signature of extended emission though the morphology is more compact.
Figure~4 shows a close-up of the north region 
showing these three radio sources. 
The compact sources VLAN~4 and VLAN~5 
were not included in this close-up, because they do not exhibit 
significant extended emission and lie far away from the main group 
($>$10${''}$). Figure ~5 shows the sources VLAN~4 and VLAN~5. 
As for the four fainter compact radio continuum sources in the
vicinity of the UC HII (VLA~2 to 5), we also looked for any 
counterpart of these other sources to gain understanding about their 
nature. The search in the Strasbourg Astronomical Data Center (CDS) and the 
Spitzer 3.6 $\mu$m band image does not show any 
counterpart within one arcsec.
We also searched for 1.4~GHz radio emission using the NRAO VLA Sky 
Survey (NVSS), which does not show any source above $\sim$1.2 mJy 
(3$\sigma$). 
The morphology of VLAN~1 to 3 resembles that of a cluster of 
radio galaxies, but the extended emission could alternatively
be HII regions 
surrounding massive stars. Such varied morphologies of HII regions are 
not uncommon (e.g Kurtz et al. 1994; 1999). However, more
observations are needed in order to clarify the nature of these 
sources. The main parameters of the five compact radio sources to
the north of G78.4+2.6 are listed in Table 2.\\

\begin{table}[htbp]\centering
\small
  \setlength{\tabnotewidth}{1.0\columnwidth} 
  \tablecols{4}
  \caption{Parameters of the radio sources toward the north of G78.4+2.6}
 \begin{tabular}{c c c c }
 \toprule
 Source &$\alpha$(2000) & $\delta$(2000)& Flux Density\tabnotemark{a}\\
  & & &(mJy)\\
 \midrule
 VLAN~1 &20 19 33.457 &40 58 53.25 &0.57$\pm$0.05\\
 VLAN~2 &20 19 36.554 &40 58 34.86 &0.87$\pm$0.04\\
 VLAN~3 &20 19 36.874 &40 58 46.14 &0.49$\pm$0.04\\
 VLAN~4 &20 19 37.997 &40 59 21.84 &0.09$\pm$0.03\\
 VLAN~5 &20 19 42.030 &40 59 22.97 &0.28$\pm$0.05\\
\bottomrule
\tabnotetext{a}{Flux densities at 3.6~cm corrected for the primary beam 
response.}
 \label{tab:2}
 \end{tabular}
\end{table}

\begin{figure*}
\centering
\includegraphics[scale=0.70, angle=0]{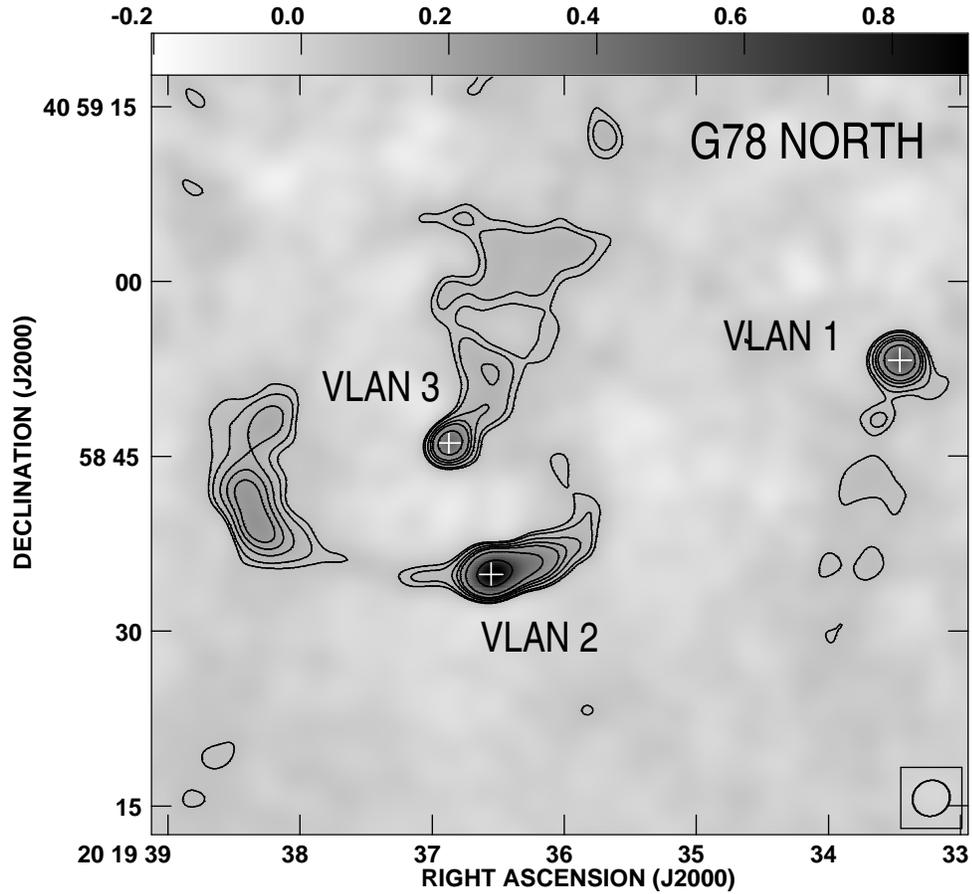}
\caption{Contour image overlapped on a grayscale image at 3.6~cm of VLAN~1, 
VLAN~2 and VLAN~3 toward the G78-North region. Contour levels 
are 3, 4, 6, 8, 10, 15, 
30 and 50 times 21.8~$\mu$Jy~beam$^{-1}$, the average rms noise of the image. 
The grayscale flux goes from -0.2 to 0.9 ~mJy~beam$^{-1}$. The synthesized 
beam, after applying UVTAPER=50 k$\lambda$ ($3\rlap.{''}26 \times 3\rlap.{''}02$
with a position angle of $-$60$^\circ$), is shown in the bottom right corner
of the image. The crosses mark the peak position of the 3.6 cm radio sources 
listed in Table 2.
}
\label{fig4}
\end{figure*}

\begin{figure}
\centering
\includegraphics[scale=0.43, angle=0]{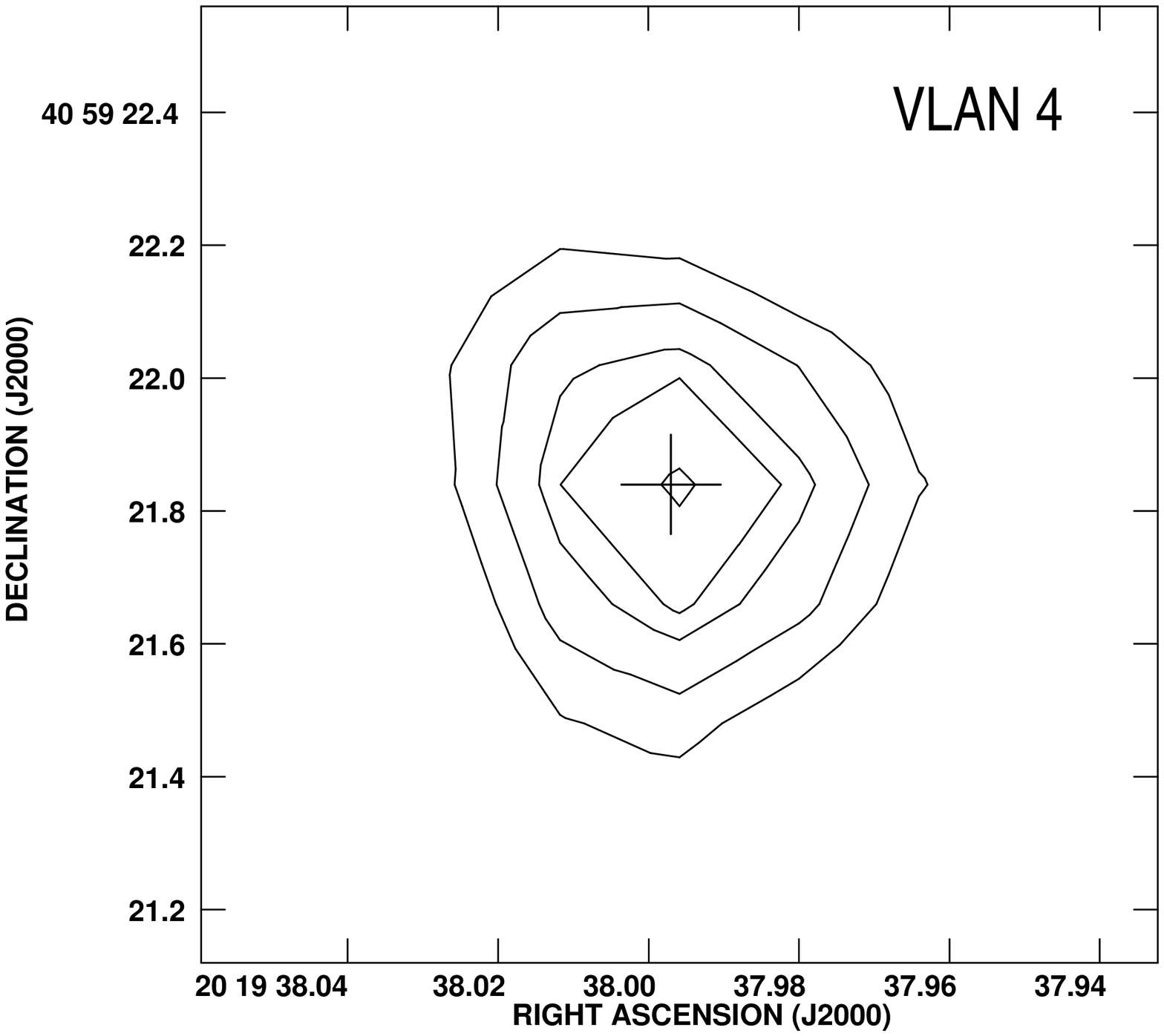}
\includegraphics[scale=0.43, angle=0]{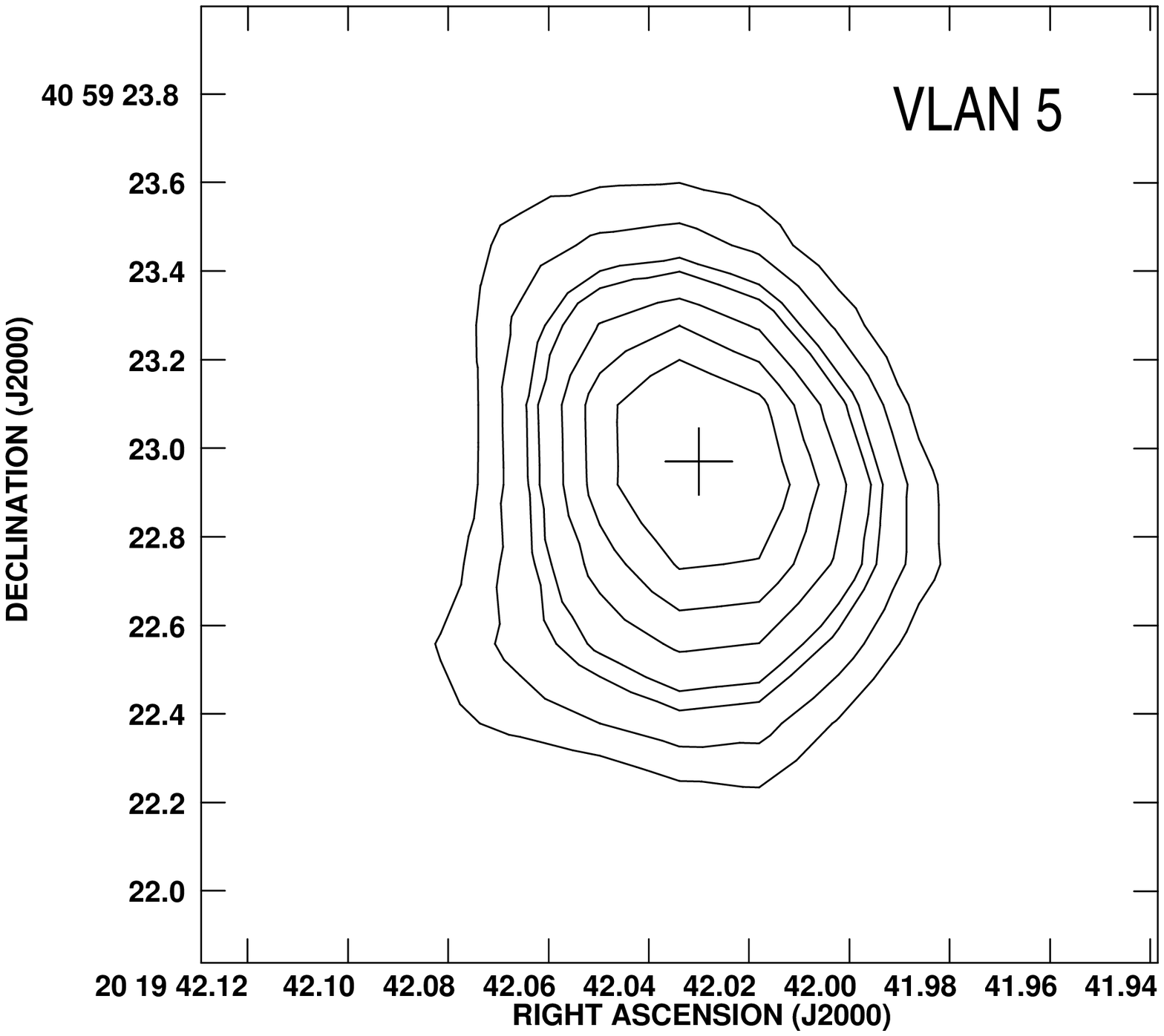}
\caption{Contour image of the 3.6 cm continuum
emission from VLAN~4 (top) and VLAN~5 (bottom) toward the 
G78-North region. The contours are -5, 5, 7, 9, 10, 12, 14 and 16 
times 10~$\mu$Jy~beam$^{-1}$, the average rms noise of the image.
The synthesized beam is 
$0\rlap.{''}76 \times 0\rlap.{''}67$, with a position angle of $-$79$^\circ$.
The crosses mark the peak position of the 3.6 cm radio sources listed in
Table 2. 
}
\label{fig5}
\end{figure}

\section{Discussion}
The high sensitive continuum observations toward the 
region G78.4+2.6 (IRAS~20178+4046), have resulted in
the identification of four new compact radio sources in the vicinity
($<$1') of the cometary HII region (VLA~1) (see Table~1).
These sources could be associated with pre-main sequence (PMS) stars,
but multiwavelength observations are required to confirm this.

Tej et al. (2007), based on the analysis of the 850 $\mu$m emission, 
have reported toward this star-forming region the presence of two 
regions at different evolutionary stages, the northern and southern
cores. The southern core traces a later stage of 
evolution and is associated with the UC HII region,
and an earlier stage of evolution is proposed for the northern core,
which does not have radio or NIR emission. 
Tej et al. (2007) have analyzed the nature of the near-infrared
sources detected in the 2MASS JHK observations, around the UC HII 
G78.4+2.6 region and by using color-magnitude diagrams and assuming
a distance of 3.3~kpc (Kurtz et al. 1994) they 
proposed the presence of nine B0.5 or earlier type candidates, all mainly 
located toward the southern core. 
However, only one (VLA~1) of the five radio sources, 
reported by us 
in this paper (Table~1), clearly coincides in position with one of the 
nine B0.5 or earlier type sources proposed by Tej et al. (2007).

The UC HII region (VLA~1) is excited at least by a B0.5 ZAMS star  (see
discussion below) and
radio emission is clearly detected. We believe that if more B0.5
stars were present in the surroundings they should have been detected in
our deep radio image. 
The non detection of the rest of the NIR sources could be due to 
a factor already considered by Tej et al. (2007): there is considerable
uncertainty in the distance to G78.4+2. Indeed, if instead of the large
distance value of 3.3~kpc we consider the smaller distance 
value of 1.7 kpc (Wilking et al. 1989; Schneider et al. 2006) then the 
B0.5 candidates will become 
less luminous stars, with no expected detectable HII region.

In the particular case of VLA~4, if it is associated with the 
NIR source number 1 of Tej et al., and the radio continuum 
reported in this work
at 3.6~cm is produced by an optically
thin homogeneous spherical HII region, following Mezger \& Henderson (1967), 
we can estimate the number of ionizing photons per second, N$_i$, produced by 
the exciting star as,
$$\bigg[\frac{N_i}{s^{-1}}\bigg] = 7.7~\times 10^{43}
\bigg[\frac{S_\nu}{mJy}\bigg]\bigg[\frac{\nu}{GHz}\bigg]^{0.1}
\bigg[\frac{D}{kpc}\bigg]^2,$$
where S$_\nu$ is the flux density of the compact source,
$\nu$ is the observing frequency and D is the distance from
the Sun. Adopting a distance of 1.7~kpc, we estimate for VLA~4 
N$_i$=1.3$\times$10$^{44}~s^{-1}$,
implying an exciting star type B3 ZAMS (Panagia 1973), 
while for a distance of 3.3~kpc, N$_i$=5.0$\times$10$^{44}~s^{-1}$,
then the spectral type will be $\sim$ B2 ZAMS.
In both cases the exciting star for the VLA~4 source should be
later than a B2 ZAMS and does not reach the ionizing rate of
a B0.5 ZAMS or earlier star.
As mentioned above, Tej et al. (2007) had already pointed out 
that there is a discrepancy seen in their determination of the 
spectral types by using the NIR 
color-magnitude diagrams due to the distance uncertainty, 
possible presence of IR excess and 2MASS resolution limit.

It is possible to plot the (H-K) colors and K magnitudes from our
VLA sources with NIR counterparts in a color-magnitude diagram to 
derive the spectral
types of the sources in the same way as done by Tej et al. (2007).
Assuming that VLA~1 is ionized by the NIR source number 2
of Tej et al. (2007), a B0.5 ZAMS star at a distance
of 1.7 kpc is needed to provide the ionization. Then, if we apply a similar
distance correction in the color-magnitude
diagram (Figure 3 of Tej et al. 2007) for all the other sources,
we find that VLA~4 (source number 1
of Tej) is associated with a B3 ZAMS star (consistent
with our previous estimate from the radio continuum
flux density). VLA~2 and VLA~3 will
correspond to a spectral type between F and G ZAMS.
Only one source shows clearly evidence of fast (hours) radio 
variability (VLA~5) a physical characteristic of low-mass PMS stars.
Unfortunately, this source does not have a NIR counterpart,
although it does have a Spitzer (mid-infrared)
counterpart (see Table~1).
We then propose that VLA~1 is an HII region
excited by a B0.5 ZAMS star, VLA~4 is possibly
an HII region excited by a B3 ZAMS star
and that the remaining three compact radio sources (VLA~2, VLA~3
and VLA~5) are low-mass PMS stars embedded in the star forming region.

The non detection in our data of radio continuum counterparts 
to the NIR sources number 3 to 9 from Tej et al. (2007), makes us 
favor the smaller distance value of 1.7 kpc toward G78.4+2.6.
Also we notice that four of our radio sources are located toward 
the southern core, where radio and 3.6 $\mu$m emission is present 
(see Figure~2), suggesting their association with a later 
evolutionary stage core where massive stars have formed an UC HII region.

The compact radio source VLA~3 is displaced $\sim$16${''}$ from 
the center of the northern core. The northern core is
traced by the submillimeter emission 
at 850 $\mu$m (see Figure~5 of Tej et al. 2007) and it is not
associated with radio or NIR emission suggesting an early stage
of evolution (Klein et al. 2005; Tej et al. 2006, 2007). We
propose that VLA~3 may be a PMS star embedded in the early 
evolutionary stage northern core.
Alternatively, VLA~3 could simply be an outer component of
the southern core that appears
projected in the line-of-sight toward the northern core. If the
association with the northern core is confirmed, it will support
the idea that the less massive objects are among the first ones to form
in a cluster (Klein et al. 2005; Molinari et al. 2008). 

In order to be sure that the four compact radio sources identified
in this paper are new members of the PMS cluster toward G78.4+2.6 
and not background sources, we now estimate the number count 
probability following Windhorst et al. (1993). 
The expected number of background sources per square arcmin, N,  
with a flux density at 8.3 GHz above S, is given by,
$$\bigg[\frac{N}{(arcmin)^2}\bigg] = 0.0024\bigg[\frac{S}{mJy}\bigg]^{-1.3}.$$
Using S=0.05 mJy (5$\sigma$ of the radio continuum image), 
within a solid angle of 1$\rlap.{'}$5$\times$1$\rlap.{'}$5 we obtain 
that the expected 
number of background sources is $\sim$0.3, suggesting that the 
four compact radio sources are members of the star forming region. 

Finally, following the formulation of Mezger \& Henderson (1967) and
Rodr\'\i guez et al. (1980) for a homogeneous and spherically
symmetric region, we estimate the physical parameters of the UC HII
region G78.4+2.6, assuming that the gas has constant electron density, an
electron temperature of 8000~K (Tej et al. 2007) and a kinematic
distance of 1.7~kpc. 
At 3.6~cm we measure a total flux density of $\sim$62.3~mJy
(see Table~1), and an angular size of 4$\rlap.{''}$5$\times$3$\rlap.{''}$5 
($\sim$0.03~pc at a distance of 1.7~kpc). 
This flux density is in agreement with other radio estimates
(Wilking et al. 1989; McCutcheon 1991; Kurtz, et al. 1994; Tej et al. 2007)
in the 0.6-8.5~GHz frequency range, which is consistent with an optically
thin region.
Under these conditions we 
estimate an electron density 
$n_{e}$= 6.4~$\times$~10$^{3}$~cm$^{-3}$, 
an ionized mass $M_{HII}$= 8.5~$\times$~10$^{-3}$~M$_{\hbox{$\odot$}}$, 
and an emission measure $EM$ of 1.3~$\times$~10$^{6}$~cm$^{-6}$~pc. 
Using the ionizing rates of Panagia (1973) and that the logarithm of the 
ionizing rate has an approximate linear behavior with the spectral subtype
(Vacca et al. 1996), we estimate for the ionizing star of VLA~1 a spectral 
type of B0.5 ZAMS (for the distance of 1.7~kpc) or of B0.3 ZAMS (for 
the distance of 3.3 kpc). 
The latter result is in agreement with Kurtz et al. (1994) and Tej et al. (2007)
that estimated a spectral type for the exciting source of VLA~1 
to be between B0 and B0.5 (assuming a distance of 3.3 kpc).
\\

\section{Conclusions}

We presented high angular resolution VLA observations at 3.6~cm 
toward the star-forming region G78.4+2.6. In addition to the cometary UC HII
region VLA 1, a group of four compact radio sources was found 
(VLA~2 to 5).
One of them (VLA~5) shows evidence of radio variability in a timescale of
hours. The four new compact radio sources, reported by us, coincide
in position with Spitzer 3.6 $\mu$m peaks.
From the five radio sources reported at 3.6~cm one is clearly associated
with the NIR source number 2 of Tej et al. (2007) with a spectral type B0.5.
Another VLA source (VLA~4) seems to be associated with the NIR source number
1 of Tej et al. (2007) and corresponds to a spectral type B3 ZAMS star.
The rest of the radio sources (VLA~2, VLA~3 and VLA~5) are probably PMS
stars with spectral types between F and G ZAMS. 
Our results favor the smaller distance of 1.7~kpc (Wilking et al. 1989;
Schneider et al 2006) proposed for the star
forming region G78.4+2.6.
 
Another group of five radio sources is found about $3'$ to the NW of the
HII region G78.4+2.6 whose nature is still undetermined.
\\


\acknowledgments
We thank an anonymous referee for the comments and suggestions that 
helped to improve this paper. 
CN, YG, and LR are thankful for the support
of DGAPA, UNAM, and of CONACyT (M\'exico).
This research has made use of the SIMBAD database, 
operated at CDS, Strasbourg, France.
This publication has also  made use of the Spitzer Space Telescope and
NASA/ IPAC Infrared
Science Archive, which is operated by the Jet Propulsion Laboratory,
California Institute of Technology, under contract with the National
Aeronautics and Space Administration.
\\
\vskip0.5cm

\centerline{APPENDIX}

\begin{appendix}

\section{Measuring the Flux Density as a Function of Time in the
\lowercase{\sl (u,v) \rm} Plane}

In some cases, the observer is interested in determining the
flux density of a source in the field as a function of time.
This can be achieved by making images from data on small time
intervals, but 
it is faster (since no images have to be made) to explore in the $(u,v)$ plane
different time intervals to search for variability at
different timescales.
As it is the case in the image plane $(x,y)$, it is possible to isolate the
emission from a source also in the $(u,v)$ plane, as described
below.

The visibility $V(u,v)$ for a given point $(u,v)$ is 
(Thompson et al. 1986):

$$V(u,v) = \int_{-\infty}^{+\infty} \int_{-\infty}^{+\infty}
I(x,y)~ exp[i 2 \pi (ux+vy)]~dx~dy, \eqno(1)$$

\noindent where $V(u,v)$ is the visibility, in general a complex
function, $u$ and $v$ are the projections of the baseline
of a given antenna pair along the sky coordinates $x$ and $y$,
respectively, given in units of wavelengths, and $I(x,y)$
is the intensity of the source or sources as a function of the position $(x,y)$ with
respect to the phase center.
A point source located at the phase center, $(x,y) = (0,0)$, can be described by

$$I(x,y) = S~\delta^2(x,y), \eqno(2)$$

\noindent where $S$ is the flux density of the source
and $\delta^2$ is Dirac's delta function
in two dimensions. The integral of equation (1) then becomes:

$$V(u,v) = S.  \eqno(3)$$

In this case, the real, $Re(u)$, and imaginary, $Im(u)$, parts of the visibility
are given by:

$$Re(u,v) = S,  \eqno(4)$$

$$Im(u,v) = 0.  \eqno(5)$$

Finally, the real and imaginary parts are averaged over the range of values of $u$
and $v$ for which one has measurements, $-u_{m} \leq u \leq u_{m}$,
$-v_{m} \leq v \leq v_{m}$:

$$Re = {{1} \over {2 u_{m}}} {{1} \over {2 v_{m}}}
\int_{-u_{m}}^{+u_{m}} \int_{-v_{m}}^{+v_{m}} Re(u,v) du~ dv = S, \eqno(6)$$

$$Im = {{1} \over {2 u_{m}}} {{1} \over {2 v_{m}}}
\int_{-u_{m}}^{+u_{m}} \int_{-v_{m}}^{+v_{m}} Im(u,v) du~ dv = 0. \eqno(7)$$

That is, the average over $(u,v)$ of the real and imaginary components of the visibility
of a point source at the phase center are simply $S$ and $0$, respectively.

What happens if there is an additional source in the field, that is centered
at a position different from the phase
center? Assuming again that this source is pointlike and that it is
located at the point $(x,y) = (x_1,0)$, it can be
described by

$$I(x,y) = S_1~\delta^2(x - x_1,y), \eqno(8)$$

\noindent where $S_1$ is its flux density. Again using equation (1) we
have that the real, $Re_1(u,v)$, and imaginary, $Im_1(u,v)$, parts of the visibility
are in this case given by:

$$Re_1(u,v) = S_1~ cos(2 \pi u x_1),  \eqno(9)$$

$$Im_1(u,v) = S_1~ sin(2 \pi u x_1).  \eqno(10)$$

Averaging over $(u,v)$, we obtain

$$Re_1 = {{1} \over {2 u_{m}}} {{1} \over {2 v_{m}}} 
\int_{-u_{m}}^{+u_{m}} \int_{-v_{m}}^{+v_{m}} Re_1(u,v) du~dv = $$
$${{1} \over {2 u_{m}}} \int_{-u_{m}}^{+u_{m}} S_1~ cos(2 \pi u x_1) du =
S_1~ {{sin(2 \pi u_{m} x_1)} \over {2 \pi u_{m} x_1}} = $$
$$S_1~ sinc(2 \pi u_{m} x_1), \eqno(11)$$

$$Im_1 = {{1} \over {2 u_{m}}} {{1} \over {2 v_{m}}}
\int_{-u_{m}}^{+u_{m}} \int_{-v_{m}}^{+v_{m}} Im_1(u,v) du~dv = $$
$${{1} \over {2 u_{m}}} \int_{-u_{m}}^{+u_{m}} S_1~ sin(2 \pi u x_1) du = 0. \eqno(12)$$

We conclude that, in this case, the average over $(u,v)$ of the
imaginary component is $0$, as in the previous case of the source
at the phase center. The real component
is now the flux density of the source, multiplied by a sinc function.
For a well-designed interferometer we can assume that $u_{m} \simeq
v_{m}$, and that the angular resolution of the data is given
approximately by

$$\theta \simeq {{1} \over {u_{m}}} \simeq {{1} \over {v_{m}}}. \eqno(13)$$

We then conclude that as long as this source is several resolution elements
away from the phase center and is not very bright, its 
contribution to the real component is small.
In brief, as long as the condition

$$S >> {{S_1} \over {2 \pi (x_1/\theta)}}, \eqno(14)$$

\noindent is fulfilled, we can take the real component of the averaged visibility 
as dominated by the flux density of the source at the phase center.

In the case of the source VLA~5, we have that $S \simeq 0.1 - 0.3$ mJy.
The other compact sources in the field of G78.4+2.6 have flux densities
comparable to that of VLA~5 and are at angular distances of $15''$ or more
away from it. Since the angular resolution of the data is $\theta \simeq 0\rlap.{''}7$, 
we have that the angular separation is $x_1 \simeq ~20~ \theta \simeq 20/u_{m}$,
and that an upper limit to the contribution of these sources is given by

$$\vert Re_1 \vert \leq 0.003~mJy. \eqno(15)$$

So, the other point sources in the field contribute less than
1\% to the real component measured after centering the \sl(u,v) \rm data
at the position of VLA~5. 

We have, however, to consider the effect of the extended source VLA~1,
that is quite bright, $S_1 \simeq 60$ mJy.
In this case it can be shown that the condition given by equation (14)
approximately becomes

$$S >> \Biggl[{{\theta} \over {\theta_S}}\Biggr]^2~ {{S_1} \over {2 \pi (x_1/\theta)}}, \eqno(16)$$

\noindent where $S_1$ is the flux density of the extended source, 
and $\theta_S$ is its angular
diameter, $\theta_S \simeq 4''$ in the case
of VLA~1. In this case $x_1 \simeq 20''$.
Evaluating the contribution of VLA~1 at the position of VLA~5 we obtain

$$\vert Re_1 \vert \leq ~ 0.01~mJy, \eqno(17)$$

$$Im_1 = 0. \eqno(18)$$

We then conclude that the contribution of VLA~1 at the position of VLA~5 does
not affect significantly the flux density determinations of the latter source.
The contribution of extended sources in the field is further suppressed
by not including in the averaging the short baselines of the \sl (u,v) \rm
data, where most of the extended emission is present.
\end{appendix}

\end{document}